\documentclass[10pt,a4paper]{article}

\usepackage{a4wide}
\usepackage{calc}
\usepackage[dvips]{graphicx}
\usepackage{amsmath}
\usepackage{amsfonts,amssymb}
\usepackage{mybbold}    

\usepackage{latexsym}           
\usepackage{multicol}
\usepackage{def}        

\usepackage[T1]{fontenc} 
\usepackage{ae}         

\usepackage{psfrag}     

\usepackage[english]{babel}

\usepackage[a4paper]{typearea}      
  \areaset[0cm]
          {17.5cm}{24cm}           

\begin{document}

\noindent
ULM-TP/02-5\\
July 2002

\newlength{\myindent}
\setlength{\myindent}{\parindent}
\vspace*{-1mm}
\begin{center}
{\Large\bf Torus quantization for spinning particles}\\[2ex]

{\bf Stefan Keppeler}\\[1ex]
{\it Abteilung Theoretische Physik, 
Universit\"at Ulm, 
Albert-Einstein-Alle 11,
D-89069 Ulm, Germany}\\
{\tt stefan.keppeler@physik.uni-ulm.de}\\[2ex]

\parbox{15cm}{\setlength{\parindent}{\myindent} \indent 
We derive semiclassical quantization conditions for systems with spin. 
To this end one has to  define the notion of integrability for the 
corresponding classical system which is given by a combination of the 
translational motion and classical spin precession. 
We determine the geometry of the invariant manifolds of this product 
dynamics which support semiclassical solutions of the wave equation. 
The semiclassical quantization conditions contain a new term, which is 
of the same order as the Maslov correction. This term is identified
as a rotation angle for a classical spin vector. Applied to the relativistic
Kepler problem the procedure sheds some light on the amazing success of 
Sommerfeld's theory of fine structure 
[Ann. Phys. (Leipzig) {\bf 51} (1916) 1--94].} 
\end{center}


\begin{multicols}{2}

Semiclassical methods for multi-component wave equations have been a topic
of constant interest over the last decade, both for their physical 
applications and the mathematical structures behind them
\cite{LitFly91a,LitFly91b,EmmWei96,BolKep98}.
In a seminal article Littlejohn and Flynn \cite{LitFly91b} summarized some 
of the previous efforts in this direction, stressed the importance of
geometric or Berry phases in this context and developed a general 
quantization scheme. Their method, however, does not cover situations in 
which the so-called principal Weyl symbol of the Hamiltonian has (globally) 
degenerate eigenvalues. But this problem shows up for the Dirac equation, 
as we will explain below. 
It was emphasized by Emmrich and Weinstein \cite{EmmWei96}
that in such a situation integrability of the so-called ray Hamiltonians 
(which in our case will be given by $H^+$ and $H^-$ defined in 
eq.~(\ref{H+-}) below)
is not a sufficient condition that allows for an explicit semiclassical 
quantization. We discuss this problem for the particular case of the Dirac 
equation, but our method also translates to more general situations.

The semiclassical analysis of the Dirac equation was started by Pauli 
\cite{Pau32} who showed that the rapidly oscillating phase of a WKB-like 
ansatz has to solve a relativistic Hamilton-Jacobi equation. 
Later Rubinow and Keller \cite{RubKel63} related the amplitude of 
the semiclassical solution to classical spin precession 
(i.e. Thomas precession \cite{Tho27}). 
So far, however, all these efforts did not result in general semiclassical 
quantization conditions as they were put forward by Keller for the 
Schr\"odinger  equation \cite{Kel58}. 
In this article we present the main steps in the derivation of 
such conditions, finally leading to eq.~(\ref{quant_cond}) below. 
To this end it will be necessary to extend the notion of integrability, see
\cite{Arn78}, from Hamiltonian systems to a certain skew product flow 
which arises naturally in the semiclassical treatment of the Dirac equation.
We also illustrate our method for an example, namely the relativistic Kepler
problem, which yields Sommerfeld's fine structure formula.

We first briefly summarize the determination of semiclassical wave functions 
for the Dirac equation. Details can be found in 
\cite{RubKel63,BolKep98,BolKep99a}. Consider the stationary Dirac equation
$\op{H}_{\rm D} \Psi = E \Psi$ with Hamiltonian 
\begin{equation}
  \op{H}_{\rm D} = 
  c\vecalph\left(\frac{\hbar}{\ui} \nabla - \frac{e}{c} \vecA(\vecx) \right) 
  + \beta mc^2 + e\phi(\vecx) 
\end{equation}
defined on a suitable domain in $L^2(\R^3) \otimes \C^4$. It describes the 
motion of a particle with mass $m$, charge $e$ and spin $\frac{1}{2}$
in electro-magnetic potentials $\phi$ and $\vecA$. The Dirac algebra is 
realized by the $4\times4$ matrices  
\begin{equation}
  \vecalph = \begin{pmatrix} 0 & \vecsig \\ \vecsig & 0 \end{pmatrix}
  \quad \text{and} \quad
  \beta = \begin{pmatrix} \eins_2 & 0 \\ 0 & -\eins_2 \end{pmatrix} \, ,
\end{equation}
where $\vecsig$ is the vector of Pauli matrices and $\eins_2$ denotes the 
$2\times2$ unit matrix. We make a semiclassical ansatz of the form 
\begin{equation}
\label{scAnsatz}
  \Psi(\vecx) 
  = \left( \sum_{k\geq0} \left(\frac{\hbar}{\ui}\right)^k a_k(\vecx) \right)
         \ue^{\frac{\ui}{\hbar}S(\vecx) }
\end{equation}
with a scalar phase function $S$ and spinor-valued amplitudes $a_k$. 
Inserting this ansatz into the Dirac equation and sorting by orders of 
$\hbar$ in leading order one finds
\begin{equation}
\label{eq:leading_order}
  \left[ H_{\rm D}(\nabla S, \vecx) - E \right] a_0 = 0
\end{equation}
with the matrix-valued function 
\begin{equation}
  H_{\rm D}(\vecp,\vecx) 
  = c\vecalph\left( \vecp - \frac{e}{c} \vecA(\vecx) \right)
    + \beta mc^2 + e\phi(\vecx) \, ,
\end{equation}
on classical phase space.
The system (\ref{eq:leading_order}) of linear equations 
only has a solution with non-trivial $a_0$ if the expression 
in square brackets has an eigenvalue zero, i.e. if $S$ solves one of the 
two Hamilton-Jacobi equations $H^\pm(\nabla S,\vecx) = E$ with classical 
Hamiltonians 
\begin{equation}
\label{H+-}
  H^\pm(\vecp,\vecx) = e\phi \pm  
    \sqrt{ c^2 \left(\vecp - \frac{e}{c} \vecA \right)^2 + m^2c^4} \,  
\end{equation}
for particles with positive and negative kinetic energy, respectively.
From standard Hamilton-Jacobi theory, see e.g. \cite{Arn78}, we conclude that
the rapidly oscillating phase of the wave function (\ref{scAnsatz}) can 
be determined by integration along solutions $(\vecP_\pm(t),\vecX_\pm(t))$ of 
Hamilton's equations of motion generated by the Hamiltonians (\ref{H+-}). 
Locally we have 
$\vecP_\pm(t) = \nabla S^\pm(\vecX_\pm(t))$,
and thus 
\begin{equation}
\label{S_als_Wegintegral}
  S^\pm(\vecx) 
  = S^\pm(\vecy) + \int_{\vecy}^{\vecx} \vecP_\pm \, \ud \vecX_\pm 
\end{equation}
where we denote by $\vecy = \vecX_\pm(0)$ the (arbitrarily chosen) starting 
point of integration. If we set $\vecxi := \vecP_\pm(0)$ we can also write
$(\vecP_\pm(t),\vecX_\pm(t)) = \phi_{H^\pm}^t(\vecxi,\vecy)$ 
with the Hamiltonian 
flows $\phi_{H^\pm}^t$. The eigenspaces corresponding to the eigenvalues
$H^\pm(\vecp,\vecx)$ of $H_D(\vecp,\vecx)$
have dimension two and we denote by $V_\pm(\vecp,\vecx)$ the $4\times2$ 
matrices of orthonormal eigenvectors, i.e. 
$V_+^\dag V_+ = \eins_2 = V_-^\dag V_-$, 
$V_+^\dag V_- = 0 = V_-^\dag V_+$ and $V_+ V_+^\dag + V_- V_-^\dag = \eins_4$,
see \cite{BolKep99a} for details. For concreteness we now seek a semiclassical 
wave function corresponding to the classical dynamics with $H^+$, and in order
to simplify notation drop the index ``$+$''. Since 
eq.~(\ref{eq:leading_order}) is a matrix equation it does not only require $S$
to solve the Hamilton-Jacobi equation, but also $a_0$ to be of the
form $a_0(\vecx) = V(\nabla S,\vecx) \, b(\vecx)$ with a $\C^2$-valued $b$.

An equation for $b$ can be derived from 
the next-to-leading order equation, obtained when inserting the
semiclassical ansatz (\ref{scAnsatz}) into the Dirac equation,
by multiplication with $V_+^\dag$ from the left, 
cf. \cite{RubKel63,BolKep98,BolKep99a}, 
\begin{align}   
\begin{split}\label{transportb}
  (\nab{p}H) \nab{x} b 
  &+ \frac{\ui}{2} \vecsig \vcB(\nab{x} S,\vecx) b\\ 
  &+ \frac{1}{2} \nab{x}[ \nab{p}H(\nab{x} S,\vecx)] b= 0 \, , 
\end{split}\\
  \vcB(\vecp,\vecx) :&= \frac{ec^2}{\varepsilon(\varepsilon+mc^2)}
  \left( \vecp-\frac{e}{c}\vecA \right) \times \vecE 
  - \frac{ec}{\varepsilon} \vecB \, . 
\end{align}     
Here we used the abbreviation $\varepsilon := \sqrt{(c\vecp-e\vecA)^2+m^2c^4}$,
and $\vecE(\vecx) = -\nabla\phi(\vecx)$ and 
$\vecB(\vecx)=\nabla\times\vecA(\vecx)$
denote the electric and magnetic fields, respectively.
Viewed as an equation along the orbit $\phi_H^t(\vecxi,\vecy)$,
the first term in (\ref{transportb})
constitutes a time derivative along the classical translational
dynamics which we shall denote by a dot. The solution of
(\ref{transportb}) with vanishing $\vcB$ is known to be given by 
$\sqrt{\det\frac{\partial\vecy}{\partial\vecx}}$, see e.g. \cite{Kel58}, 
and thus the ansatz $b = \sqrt{\det\frac{\partial\vecy}{\partial\vecx}} \, u$
leaves us with the spin transport equation 
\begin{equation}
\label{spintransport}
  \dot{u} + \frac{\ui}{2} \vecsig \vcB(\phi_H^t(\vecxi,\vecy)) \, u 
  = 0 \, . 
\end{equation}
The solution of (\ref{spintransport}) can be written as 
$u(t) = d(\vecxi,\vecy,t) u(0)$ with an $\SU(2)$-matrix $d(\vecxi,\vecy,t)$.
We explicitly indicate the dependence on the initial point 
$(\vecxi,\vecy)$ of the classical trajectory along which 
we integrate until time $t$. 
Through the covering map $\varphi: \SU(2) \to \SO(3)$
we can associate with the spin transporter $d$ a rotation matrix 
$R(\vecxi,\vecy,t)$, and one easily verifies that 
$\vecs(t) := R(\vecxi,\vecy,t) \, \vecs(0)$ 
solves the spin precession equation
\begin{equation}
\label{spinprec}
  \dot{\vecs} = \vcB(\phi_H^t(\vecxi,\vecy)) \times \vecs  
\end{equation}
on the two-sphere $S^2$ (i.e. $\vecs \in \R^3$, $|\vecs|=1$).
This is the equation of Thomas precession \cite{Tho27} thus emerging
from a semiclassical analysis of the Dirac equation.
It turns out that all properties of the semiclassical wave function 
$\Psi \sim a_0 \exp(\frac{\ui}{\hbar}S)$ can be determined from the solution 
$\phi_H^t(\vecxi,\vecy)$ of Hamilton's equations of motion and the solution 
$\vecs(t)$ of eq.~(\ref{spinprec}).
Thus the skew product 
\begin{equation}
\label{Ycl}
  Y_{\rm cl}^t (\vecxi,\vecy,\vecs(0)) 
  := \left( \phi_{H}^t(\vecxi,\vecy), R(\vecxi,\vecy,t) \vecs(0) \right) ,
\end{equation}
which defines a flow on the extended classical phase space 
$\R^{2d} \times S^2$, should be considered as the classical dynamical 
system corresponding to the Dirac equation, cf. \cite{BolKep99b,BolGlaKep01}. 

The key question in semiclassical quantization is now whether 
it is possible to find a single-valued wave function 
$\Psi \sim a_0 \exp(\frac{\ui}{\hbar}S)$ which solves the above 
equations. Let us briefly recall the procedure in the spinless case 
\cite{Kel58}.

In standard semiclassics for the Schr\"odinger equation 
one invokes integrability of the classical flow $\phi_H^t$: Besides 
the classical Hamiltonian $H=:A_1$ there are $d-1$ further conserved 
quantities, $A_2, \hdots, A_d$ (for a system with $d$ degrees of freedom; 
we only specialize to $d=3$ later) with 
mutually vanishing Poisson brackets, $\{A_j,A_k\}=0$. Then the Theorem 
of Liouville and Arnold, see \cite[chapter 10]{Arn78}, guarantees
that a (compact and connected) invariant level set 
$\{(\vecp,\vecx) \, | \, \vecA = const. \}$ 
has the topology of a $d$-torus $\T^d$ on which the flows 
$\phi_{A_1}^t, \hdots, \phi_{A_d}^t$ generated by $A_1,\hdots,A_d$
commute. By integration along the flow 
lines of $\phi_{A_2}^t, \hdots, \phi_{A_d}^t$ -- analogous to the integration 
along $\phi_H^t$ in (\ref{S_als_Wegintegral}) -- this allows for a definition
of the phase function $S$ which is unique up to the contributions of
non-contractible loops. Demanding single-valuedness of the semiclassical 
wave function $\Psi \sim a_0 \ue^{\frac{\ui}{\hbar}S}$ yields the 
Einstein-Brillouin-Keller (EBK) quantization conditions 
\begin{equation}
\label{EBK}
  \oint_{\mathcal{C}_j} \vecp \, \ud\vecx 
  = 2\pi\hbar \left( n_j + \frac{\mu_j}{4} \right) \, , \quad n_j \in \Z \, , 
\end{equation}
where $\{ \mathcal{C}_j \, | \, j=1,\hdots,d \}$ denotes a basis of 
non-contractible loops on the torus characterized by the action variables 
$I_j = \frac{1}{2\pi} \oint_{\mathcal{C}_j} \vecp \, \ud\vecx$.
The number $\mu_j \in \{ 1,2,3,4 \}$ is the Maslov index, see \cite{MasFed81},
of the cycle $\mathcal{C}_j$ which, roughly speaking, counts the number 
of points along $\mathcal{C}_j$ at which the pre-factor 
$\sqrt{\det\frac{\partial\vecy}{\partial\vecx}}$ becomes singular. All 
these terms also appear in the situation with non-zero spin, and
we now have to examine how the spin contribution modifies this picture. 

When we include the spin contribution $d(\vecxi,\vecy,t)$ the situation 
becomes more complicated and integrability of $\phi_H^t$ will, in general, not 
be a sufficient condition to allow for an explicit semiclassical quantization. 
This can be seen as follows: Transporting the spinor-valued amplitude $u$ 
along a closed path $\mathcal{C}_j$ on a Liouville-Arnold
torus the initial and final value, $u_{\rm i}$ and $u_{\rm f}$, 
respectively, differ not only by a phase but are related by an 
$\SU(2)$-transformation, $u_{\rm f} = d_j u_{\rm i}$, $d_j \in \SU(2)$.
Mathematically speaking, we are considering a connection in a $\C^2$-bundle
with $\SU(2)$-holonomy. If there was only one such loop, as in a system 
with one translational degree of freedom, we could choose $u_{\rm i}$ 
to be an eigenvector of $d_j$, thus reducing the $\SU(2)$-holonomy to a simple
phase factor. However, for $d \geq 2$ degrees of freedom this is impossible
since the holonomy factors for different loops are, in general, given by 
non-commuting elements of the holonomy group $\SU(2)$. This is a general 
problem in semiclassics for multi-component wave equations with 
globally degenerate eigenvalues of the principal symbol, as was emphasized 
in a general setting by Emmrich and Weinstein \cite{EmmWei96}.

In our situation of semiclassics for spinning particles we will 
solve this problem by imposing additional conditions on the ``field'' $\vcB$, 
which generates the classical spin precession (\ref{spinprec}). From 
a physical point of view it is not surprising that we need a stronger 
condition than just integrability of the translational dynamics $\phi_H^t$;
since we identified the skew product (\ref{Ycl}) as the classical dynamics
corresponding to the Dirac equation, we should also say under which
circumstances  we want to call the spin dynamics (or rather the combination of 
translational and spin dynamics)  integrable. We do this by the following 
definition.\\[1ex]
{\bf Definition} {\it The skew product $Y_{\rm cl}^t$ is called integrable, 
if (i) the underlying Hamiltonian flow $\phi_H^t$ is integrable in the sense 
of Liouville and Arnold and (ii) the flows $\phi_{A_2}^t,\hdots,\phi_{A_d}^t$ 
can also be extended to skew products $\Ycl_j^t$ on 
$\R^{2d} \times S^2$ ($Y_{\rm cl}^t \equiv \Ycl_1^t$)
with fields $\vcB_j$ fulfilling 
\begin{equation}
\label{spin_involution}
  \{A_j,\vcB_k\} + \{\vcB_j,A_k\} - \vcB_j \times \vcB_k = 0 
  \, . 
\end{equation}
}%
Condition (\ref{spin_involution}) plays the same role as  
the condition $\{A_j,A_k\}=0$ does in the scalar case; 
it guarantees that all skew products $\Ycl_j^t$ commute \cite{Kep_prep}.
Under these conditions we are able to prove the following theorem.\\[1ex]
{\bf Theorem} {\it If the skew product flow $Y_{\rm cl}^t$ is integrable,
the combined phase space $\R^{2d} \times S^2$ can be
decomposed into invariant bundles 
$\mathcal{T}_\theta \stackrel{\pi}{\longrightarrow} \T^d$ over 
Liouville-Arnold tori $\T^d$ with fiber $S^1$. The bundles can be embedded
in $\T^d \times S^2$ such that the fibers are characterized by the 
latitude with respect to a local direction $\vecn(\vecp,\vecx)$, i.e.
\begin{equation}
\label{T_theta}
  \mathcal{T}_\theta = \{ (\vecp,\vecx,\vecs) \in \T^d \times S^2 \, | \,
        \sphericalangle (\vecs,\vecn(\vecp,\vecx)) = \theta \} \, . 
\end{equation}
}%
The proof of this theorem will be given elsewhere \cite{Kep_prep}. 
The geometry of the the invariant sets $\mathcal{T}_\theta$ 
is illustrated in figure $\ref{fig:torus_mit_sphaere}$: a Liouville-Arnold 
torus is sketched as a $2$-torus; at two different points we show the attached
sphere together with the local axes $\vecn$ and a corresponding parallel
of latitude.

\noindent
\hspace*{0.5\parindent}
\parbox{\columnwidth-\columnsep-\parindent}{
\vspace{5ex}
\psfrag{(p,x)}{$(\vecp,\vecx)$}
\psfrag{(p',x')}{$(\vecp^\prime,\vecx^\prime)$}
\psfrag{s}{$\vecs$}
\psfrag{s'}{$\vecs^\prime$}
\psfrag{n(p,x)}{$\vecn(\vecp,\vecx)$}
\psfrag{n(p',x')}{$\vecn(\vecp',\vecx')$}
\includegraphics[width=7.5cm]{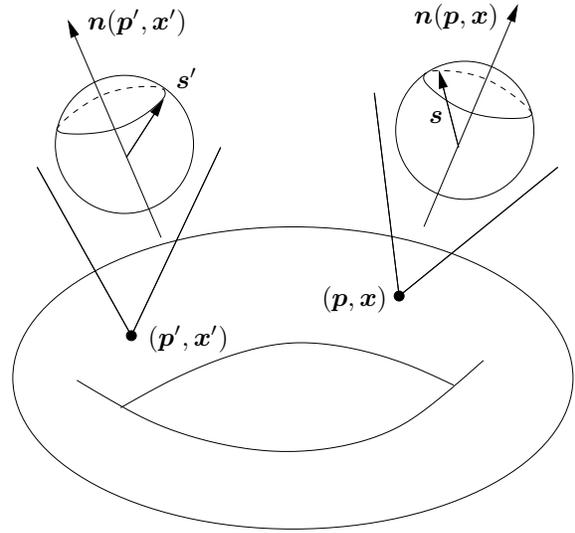}
\captionof{figure}{%
The invariant manifolds $\mathcal{T}_\theta$ of $Y_{\rm cl}^t$, 
see (\ref{T_theta}), are given by tori $\T^d$ to which at each point is 
attached the set of all points on the two-sphere $S^2$ sharing a fixed 
latitude $\theta$ 
with respect to a varying axis $\vecn(\vecp,\vecx)$.}
\label{fig:torus_mit_sphaere}
\vspace{5ex}
}

If the skew product flow $Y_{\rm cl}^t$ is integrable, the theorem
allows us to construct semiclassical wave functions which imply generalized
quantization conditions involving the spin degree of freedom. We briefly 
sketch the construction and then state the quantization conditions.

As in the case without spin we define the semiclassical wave function
by integration along the flow lines of $\phi_{A_1}^t, \hdots, \phi_{A_d}^t$.
In addition we choose the $\C^2$-valued part $u$ such that it is an 
eigenvector of $\vecsig\vecn(\vecp,\vecx)$ at each point of the 
Liouville-Arnold torus $\T^d$. 
(This is only possible if the skew product $Y_{\rm cl}^t$,
and not just the Hamiltonian flow $\phi_H^t$, is integrable .) 
Then the semiclassical wave function is unique up to the contribution of 
non-contractible loops on $\T^d$. Transporting a classical spin vector 
along such a loop $\mathcal{C}_j$ by a combination of the (commuting)
skew products $\Ycl_1^t, \hdots, \Ycl_d^t$, one finds that it is rotated by an 
angle $\alpha_j$, while integrability of $Y_{\rm cl}^t$ ensures that 
it stays on the same parallel of latitude. Consequently, the semiclassical
wave function is multiplied by a phase factor $\ue^{\mp\ui\alpha_j/2}$, the
sign depending on whether we have chosen $u$ to be an eigenvector of 
$\vecsig\vecn$ with eigenvalue $+1$ or $-1$. Demanding single-valuedness of
the wave function, the total phase change when moving along a loop
$\mathcal{C}_j$ has to be an integer multiple of $2\pi$, yielding the 
quantization conditions
\begin{equation}
\label{quant_cond}
  \oint_{\mathcal{C}_j} \vecp \, \ud \vecx = 
  2\pi \hbar \left( n_j + \frac{\mu_j}{4} + m_s \frac{\alpha_j}{2\pi} \right)
  \, , 
\end{equation}
where in addition to the terms in (\ref{EBK}) the spin contribution 
with the spin quantum number $m_s = \pm \frac{1}{2}$ enters.

We remark that analogous quantization conditions can be derived for the 
Pauli equation \cite{Kep_prep}. 
There we can also choose to describe particles with arbitrary 
spin $s\in\frac{1}{2}\N_0$ by replacing the Pauli matrices $\vecsig$ with
a higher dimensional irreducible representation of $\su(2)$. This changes
neither the corresponding classical system (which is always given by 
a skew product on $\R^{2d} \times S^2$) nor the construction of the 
semiclassical solutions; only in the quantization conditions 
(\ref{quant_cond}) the spin quantum number $m_s$ then takes the values 
$-s, -s+1, \hdots, s$.

We conclude by illustrating these new quantization conditions for a famous 
example, namely Sommerfeld's fine structure formula \cite{Som16}. To this end
we have to quantize the relativistic Kepler problem with classical Hamiltonian
\begin{equation}
  H(\vecp,\vecx) = -\frac{e^2}{|\vecx|} + \sqrt{c^2\vecp^2 + m^2c^4} \, . 
\end{equation}
The problem can be transformed to action and angle variables, 
see e.g. \cite{Som16}, and the new Hamiltonian depends only on the 
two action variables $I_r$ and $L$. Here $I_r$ denotes the action variable 
corresponding to a radial cycle (e.g. from perihelion to aphelion and back),
and $L$ is the modulus of angular momentum $\vecL = \vecx \times \vecp$.
In 1916 Sommerfeld quantized this system using the old quantum theory, since 
quantum mechanics was still to be invented, not to think
about spin or the Dirac equation. Accordingly, he chose the quantization 
conditions
\begin{equation}
  I_r = \hbar n_r \quad \text{and} \quad L = \hbar l
\end{equation}
with integers $n_r \in \N_0$ and $l \in \N$. 
More than ten years later it was confirmed
that the energy levels resulting from these conditions are exactly the same
as one finds by solving the corresponding Dirac equation \cite{Gor28,Dar28}.
This is insofar surprising as the Dirac equation not only includes 
relativistic effects, but also takes into account spin-orbit coupling, 
which Sommerfeld could not know about. Quantizing the problem with the 
new conditions (\ref{quant_cond}) yields
\begin{equation}
  I_r = \hbar \left( n_r + \frac{1}{2} \pm \frac{\alpha_r}{2\pi} \right)
  \quad \text{and} \quad 
  L = \hbar \left( l + \frac{1}{2} \pm \frac{\alpha_L}{2\pi} \right)
\end{equation}
with integers $n_r$ and $l$ and a Maslov contribution of $\frac{1}{2}$ for
both variables. For the spin rotation angle $\alpha_L$ we find 
$\alpha_L = 2\pi$ for any spherically symmetric system \cite{Kep_prep}. 
Intriguingly, for the relativistic Kepler problem $\alpha_r$ is also 
given by $2\pi$! Therefore, the conditions (\ref{quant_cond}) and 
Sommerfeld's method yield the same values for $I_r$ and $L$, thus leading
to the same energy levels. A careful analysis of the values that $n_r$ and 
$l$ can assume (one finds $n_r \geq 0$ and $l\geq \frac{1}{2} \mp \frac{1}{2}$)
shows that with the semiclassical quantization scheme developed 
here one also obtains the correct multiplicities, which Sommerfeld was  
unable to extract with his method. 

Summarizing, we can say that, by a freak 
of nature, Sommerfeld was able to obtain the correct energy levels of the
Dirac hydrogen atom because, roughly speaking, the corrections due to wave 
mechanics (the Maslov term $\frac{1}{2}$) and those due to the spin 
$\frac{1}{2}$ of the electron cancel for this particular problem. 

I would like to thank Jens Bolte for helpful discussions 
and I gratefully acknowledge financial support from the 
Deutsche Forschungsgemeinschaft (DFG) under contract no. Ste 241/10-2.
%
%
\vspace*{-4ex}
%

\begin{thebibliography}{10}
\setlength{\itemsep}{-4pt}\vspace*{-2ex}
\bibitem{LitFly91a}
R.~G. Littlejohn and W.~G. Flynn: {\em {G}eometric {P}hases in the
  {B}ohr-{S}ommerfeld {Q}uantization of {M}ulticomponent {W}ave {F}ields\/},
  Phys. Rev. Lett. {\bf 66} (1991) ~2839--2842.

\bibitem{LitFly91b}
R.~G. Littlejohn and W.~G. Flynn: {\em {G}eometric phases in the asymptotic
  theory of coupled wave equations\/}, Phys. Rev. A {\bf 44} (1991)
  ~5239--5256.

\bibitem{EmmWei96}
C.~Emmrich and A.~Weinstein: {\em {G}eometry of the transport equation in
  multicomponent {WKB} approximations\/}, Commun. Math. Phys. {\bf 176} (1996)
  ~701--711.

\bibitem{BolKep98}
J.~{B}olte and S.~{K}eppeler: {\em {S}emiclassical Time Evolution and Trace
  Formula for Relativistic Spin-1/2 Particles\/}, Phys. Rev. Lett. {\bf 81}
  (1998) ~1987--1991.

\bibitem{Pau32}
W.~{P}auli: {\em {D}iracs {W}ellengleichung des {E}lektrons und geometrische
  {O}ptik\/}, {H}elv. {P}hys. {A}cta {\bf 5} (1932) ~179--199.

\bibitem{RubKel63}
S.~I. Rubinow and J.~B. Keller: {\em {A}symptotic {S}olution of the {D}irac
  {E}quation\/}, {P}hys. {R}ev. {\bf 131} (1963) ~2789--2796.

\bibitem{Tho27}
L.~H. Thomas: {\em The Kinematics of an Electron with an Axis\/}, Philos. Mag.
  {\bf 3} (1927) ~1--22.

\bibitem{Kel58}
J.~B. Keller: {\em {C}orrected {B}ohr-{S}ommerfeld {Q}uantum {C}onditions for
  {N}onseparable {S}ystems\/}, {A}nn. {P}hys. ({NY}) {\bf 4} (1958) ~180--185.

\bibitem{Arn78}
V.~I. Arnold: {\em Mathematical Methods of Classical Mechanics\/},
  {S}pringer-{V}erlag, {N}ew {Y}ork,  (1978).

\bibitem{BolKep99a}
J.~{B}olte and S.~{K}eppeler: {\em A semiclassical approach to the {D}irac
  equation\/}, Ann. Phys. (NY) {\bf 274} (1999) ~125--162.

\bibitem{BolKep99b}
J.~{B}olte and S.~{K}eppeler: {\em {S}emiclassical form factor for chaotic
  systems with spin 1/2\/}, J. Phys. A {\bf 32} (1999) ~8863--8880.

\bibitem{BolGlaKep01}
J.~Bolte, R.~Glaser and S.~Keppeler: {\em Quantum and classical ergodicity of
  spinning particles\/}, Ann. Phys. (NY) {\bf 293} (2001) ~1--14.

\bibitem{MasFed81}
V.~P. Maslov and M.~V. Fedoriuk: {\em {S}emi-{C}lassical {A}pproximation in
  {Q}uantum {M}echanics\/}, {D}. {R}eidel, {D}odrecht,  (1981).

\bibitem{Kep_prep}
S.~Keppeler: (in preparation).

\bibitem{Som16}
A.~Sommerfeld: {\em {Z}ur {Q}uantentheorie der {S}pektrallinien\/}, Ann. Phys.
  (Leipzig) {\bf 51} (1916) ~1--94, 125--167.

\bibitem{Gor28}
W.~Gordon: {\em {D}ie {E}nergieniveaus des {W}asserstoffatoms nach der
  {D}iracschen {Q}uantentheorie des {E}lektrons\/}, Z. Phys. {\bf 48} (1928)
  ~11--14.

\bibitem{Dar28}
C.~G. Darwin: {\em {T}he {W}ave {E}quations of the {E}lectron\/}, Proc. R. Soc.
  London Ser. A {\bf 118} (1928) ~654--680.

\end{thebibliography}
%

\end{multicols}

\end{document}